\documentclass[preprint,tighten,amsmath,amssymb,aps,twocolumn]{aastex701}
\usepackage{graphicx}
\usepackage{comment}
\usepackage{color}
\newcommand{\msun}{M\ensuremath{_{\odot}}}

\newcommand{\kms}{km s\ensuremath{^{-1}}}
\newcommand{\mbh}{\ensuremath{M_\mathrm{BH} }}

\begin{document}

\title{Mapping the Stellar Kinematics in the Central 240 Parsecs of M87 with the James Webb Space Telescope}

\author[0009-0008-8003-1048]{Refa M.~Al-Amri}
\affiliation{George P.~and Cynthia Woods Mitchell Institute for Fundamental Physics and Astronomy, and Department of Physics and Astronomy, Texas A\&M University, College Station, TX 77843, USA}
\email{refaalamri@tamu.edu}

\author[0000-0002-1881-5908]{Jonelle L.~Walsh}
\affiliation{George P.~and Cynthia Woods Mitchell Institute for Fundamental Physics and Astronomy, and Department of Physics and Astronomy, Texas A\&M University, College Station, TX 77843, USA}
\email{walsh@tamu.edu}

\author[0000-0002-7703-7077]{Emily R.~Liepold}
\affiliation{Department of Astronomy, University of California, Berkeley, CA 94720, USA}
\email{emilyliepold@berkeley.edu}

\author[0000-0002-4430-102X]{Chung-Pei Ma}
\affiliation{Department of Astronomy, University of California, Berkeley, CA 94720, USA}
\affiliation{Department of Physics, University of California, Berkeley, CA 94720, USA}
\email{cpma@berkeley.edu}

\author[0000-0002-5612-3427]{Jenny E.~Greene}
\affiliation{Department of Astrophysical Sciences, Princeton University, 4 Ivy Lane, Princeton, NJ 08544, USA}
\email{jgreene@astro.princeton.edu}

\correspondingauthor{Refa M.~Al-Amri}
\email{refaalamri@tamu.edu}

\begin{abstract}

The supermassive black hole (SMBH) in the giant elliptical galaxy M87 is one of the most well-studied in the local universe, but the stellar- and gas-dynamical SMBH mass measurements disagree. As this galaxy is a key anchor for the upper end of the SMBH mass$-$host galaxy relations, we revisit the central $3\arcsec\times 3\arcsec$ ($\sim 240\times240$ pc) region of M87 with the Near Infrared Spectrograph (NIRSpec) integral field unit (IFU) on the James Webb Space Telescope (JWST). We implement several improvements to the reduction pipeline and obtain high signal-to-noise spectra ($S/N \sim 150$) in single 0\farcs05 spaxels across much of the NIRSpec field of view. We measure the detailed shape of the stellar line-of-sight velocity distribution, parameterized by Gauss-Hermite moments up to $h_8$, in $\sim 2800$ spatial bins, substantially improving upon the prior high angular resolution studies of the M87 stellar kinematics. The NIRSpec data reveal velocities with $V \sim \pm 45$ \kms, velocity dispersions that rise sharply to $\sim$$420$ \kms\ at a projected radius of 0\farcs45 (36 pc), and a slight elevation in $h_4$ toward the nucleus. We comprehensively test the robustness of the kinematics, including using multiple velocity template libraries and adopting different polynomials to adjust the template spectra. We find that the NIRSpec stellar kinematics seamlessly transition to recently measured large-scale stellar kinematics from optical Keck Cosmic Web Imager (KCWI) IFU data. These combined NIRSpec and KCWI kinematics provide continuous coverage from parsec to kiloparsec scales and will critically constrain future stellar-dynamical models of M87.

\end{abstract}


\section{Introduction} \label{sec:intro}

Supermassive black holes (SMBHs) are believed to play a significant role in galaxy growth and evolution. This has been established on the basis of several relations between SMBH mass ($\mbh$) and large-scale properties of the host galaxy, such as bulge luminosity and stellar velocity dispersion \citep{Kormendy&Ho2013, McConnell&Ma2013, Saglia2016}. The relations have further broad, far-reaching implications. They inform models of black hole feeding and feedback \citep{Silk_Rees_1998, Fabian_1999, Shankar_2009}, set the SMBH mass function important for making gravitational wave estimates for Pulsar Timing Arrays and space-based detectors \citep{Agazie_2023, Matt_2023, Liepold_2024}, serve as a reference to which reverberation-mapped active galactic nuclei (AGN) are compared to determine a virial coefficient needed to estimate \mbh\ from single-epoch spectra \citep{Reines_2013, Greene_2024}, and form the baseline when searching for possible redshift evolution \citep{Bennert_2011, Zhang_2023, Pacucci_2023}.

However, the low ($\mbh \lesssim 10^6$ \msun) and high ($\mbh \gtrsim 10^9$ \msun) mass ends of the SMBH$-$host galaxy relations are not well understood. Also, the relations aren't as simple as once thought, as galaxies with different structural properties and evolutionary histories, including brightest cluster galaxies, massive cored elliptical galaxies, compact early-type galaxies, and low-mass spiral galaxies, show surprises in the scaling relations (e.g.,~\citealt{Greene_2010, McConnell_2011, Seth_2014, Walsh_2016, deNicola_2024, Liepold_2025}). Hence, more, robust \mbh\ measurements are needed at the low and high ends of the SMBH mass distribution and in a wide range of galaxies types.

The giant elliptical galaxy, M87, with a $5.37 \times 10^9\ \msun$ SMBH from stellar dynamics \citep{Liepold2023}, is a crucial anchor to the upper end of the scaling relations. The SMBH is one of the best studied in the local universe, alongside Sagittarius (Sgr) A$^*$ in the Galactic Center at the extreme opposite end of the SMBH mass distribution. Similar to Sgr A$^*$, the increase in angular resolution over the years, culminating in the impressive Event Horizon Telescope (EHT) image of the SMBH shadow \citep{M87EHT2019A}, is now paving the way for a deeper understanding of the compact object in M87 and the immediate surrounding environment. In addition to the EHT measurement, M87 has been the subject of numerous \mbh\ measurements over the past $\sim$45 years. While the data and modeling methods have improved, the stellar- and gas-dynamical \mbh\ determinations disagree by a factor of about two (e.g.,~\citealt{Harms1994, Macchetto1997, Gebhardt2011, Walsh2013, Liepold2023}).

The stellar-dynamical \mbh\ measurements for M87, in particular, have progressed from isotropic to anisotropic models, from Jeans models to orbit-based implementations, from sampling two-integral distribution functions to three integrals of motion, from modeling the SMBH and stars to additionally incorporating dark matter, and most recently from axisymmetric models to allowing for a triaxial intrinsic galaxy shape (e.g.,~\citealt{Sargent1978, Young1978, Dressler&Richstone1990, vanderMarel1994, Magorrian1998, GebhardtThomas2009, Liepold2023}). Many of these models have been fit to observations of M87's stellar kinematics from seeing-limited, ground-based telescopes and on large spatial scales. 

In contrast, obtaining high angular resolution stellar kinematics at M87's nucleus has been challenging due to the galaxy's surface brightness core. It is difficult to acquire the high signal-to-noise ($S/N$) observations needed to securely measure the detailed shape of the stellar line-of-sight velocity distribution (LOSVD) in M87's faint galaxy center. The recent Keck Cosmic Web Imager (KCWI) study \citep{Liepold2023} provided stellar kinematic measurements from a radius of $\sim 150\arcsec$ down to $\sim 1\arcsec$ from the center of M87, well within the SMBH’s gravitational sphere of influence ($\sim 5\arcsec$). For the innermost $\sim 1\arcsec$ of M87, only two high angular resolution stellar kinematic studies have been conducted -- \cite{Gebhardt2011} collected adaptive optics (AO) observations from Gemini North's Near-infrared Integral Field Spectrograph (NIFS) and \cite{Simon2024} presented AO Very Large Telescope (VLT) narrow field mode (NFM) Multi Unit Spectroscopic Explorer (MUSE) data (originally acquired by \citealt{Osorno2023}).

With the James Webb Space Telescope (JWST), we are now positioned to improve upon the limited number of small-scale stellar kinematic studies and probe the innermost region of M87 in unprecedented detail, thanks to JWST's high angular resolution, reduced background, and enhanced sensitivity. Using JWST's Near Infrared Spectrograph (NIRSpec) in integral field unit (IFU) mode, we acquire a new view of the stellar kinematics deep within the gravitational influence of the SMBH. We present these stellar kinematics and lessons learned after significant testing of the process used to extract the stellar kinematics. The NIRSpec stellar kinematics, combined with recently published wide-field IFU stellar kinematics of M87 \citep{Liepold2023} and state-of-the-art dynamical models (e.g.,~\citealt{Vasiliev_Valluri_2020, Neureiter_2021, Quenneville2021, Quenneville2022, Tahmasebzadeh_2022}) will firmly establish M87 as a benchmark galaxy anchoring the upper end of the SMBH$-$galaxy scaling relations.

The paper is structured as follows. Section \ref{sec:obs_reduction} presents the NIRSpec observations and data reduction, including modifications we made to the default data reduction pipeline. Section \ref{sec:m87cube} describes our final data cube for M87. Sections \ref{sec:masking_binning} and \ref{sec:stellar_kin} discuss the processes of spatial masking and binning and measuring the stellar kinematics, along with the steps we took to assess the robustness of the extracted kinematics. Finally, we place our results in context of past work in Section \ref{sec:discussion} and conclude in Section \ref{sec:conclusion}. Throughout the paper, we assume a distance of 16.8 Mpc (adopted by \citealt{EHT2019} and \citealt{Liepold2023}), such that 1\arcsec\ corresponds to 81.1 pc. All wavelengths are observed and are in vacuum, unless otherwise noted.

\section{NIRSpec Observations and Data Reduction} \label{sec:obs_reduction}

We observed M87 using the JWST NIRSpec IFU with the G235H grating and the F170LP filter on 31 May 2023 as part Cycle 1 GO-2228. We used the NRSIRS2 readout mode and integrated for a total of $6.6$ hours on-source using 20 groups, 1 integration, and 16 exposures dithered with the small ($0\farcs25$ extent) cycle pattern to improve sampling of the point spread function (PSF). We also obtained a LeakCal exposure at the first dither position.

The raw data was generated using the JWST Science Data Processing version \texttt{2023$\_$2a}, and we used the Calibration Reference Data System version 11.17.3, along with the context file \texttt{jwst$\_$1140.pmap}. The version of the JWST Science Calibration Pipeline used to process the data was 1.11.4. 

During Stage 1 (\texttt{calwebb\_detector1}) of the pipeline, we perform detector-level corrections like initializing the data quality (DQ) array, checking for saturation, superbias and reference pixel subtraction, correcting for non-linearity, and removing the dark current. This stage also flags any large jumps between two consecutive groups relative to other consecutive pairs of groups. We further turn snowball flagging on to correct for large cosmic-ray impacts; this setting is now on as the default for newer versions of the pipeline. The mean count rate for each pixel is determined by a ramp fitting step and ultimately a 2D count rate image for each exposure is produced. At this point, we also explored using an external package, \texttt{NSClean} \citep{Rauscher_2024}, to correct for correlated read noise. However, we find for our data of M87, read out with NRSIRS2, this correction was unnecessary and we did not implement such a step. In newer versions of the pipeline, correcting for correlated read noise is built-in, as an optional step only, and can be implemented in Stage 2.

Next, in Stage 2 (\texttt{calwebb\_spec2}), we assign the World Coordinate System information to the data, thus mapping detector pixels to sky and wavelength coordinates. We subtract the LeakCal exposure, previously processed through Stage 1 of the pipeline, from each M87 exposure to remove the imprint from the NIRSpec Micro-Shutter Assembly (MSA). We further flag regions that were affected by MSA failed open shutters, apply a flat-field correction, correct for signal loss due to aperture and optical system effects, and complete the photometric calibration. The resulting products are calibrated exposures with units of MJy/sr.

Finally, Stage 3 (\texttt{calwebb\_spec3}) of the pipeline identifies outliers that were not flagged in Stage 1 and generates a combined data cube of all calibrated exposures. However, we find that we could improve both of these steps, and we detail the changes we made relative to the default pipeline below.

\subsection{Removing Additional Outliers} \label{subsec:removing_outliers}

We complement the outlier detection step in Stage 3 of the pipeline by manually flagging remaining bad pixels and artifacts in each of the exposures, and adjusting the DQ array to `\texttt{DO\_NOT\_USE}' before assembling the data cube. With many dithered exposures, bad pixels in a single exposure can be ignored when constructing the final cube. Along similar lines, others have flagged and sometimes corrected bad pixels in the 2D detector images (e.g., \citealt{Perna_2023, Hutchison_2024, Bianchin_2024, Tahmasebzadeh_2025}). Another approach is to identify artifacts in the final 3D data cube (e.g., \citealt{Perna_2023, Bohn_2024}), but flagging bad pixels in each of the 2D calibrated detector images is preferable because any artifacts in the detector image will be spread out and made worse by the interpolation during the cube-building process.

\begin{figure}[h!]
\begin{singlespace}
  \centering
  \includegraphics[width=\linewidth]{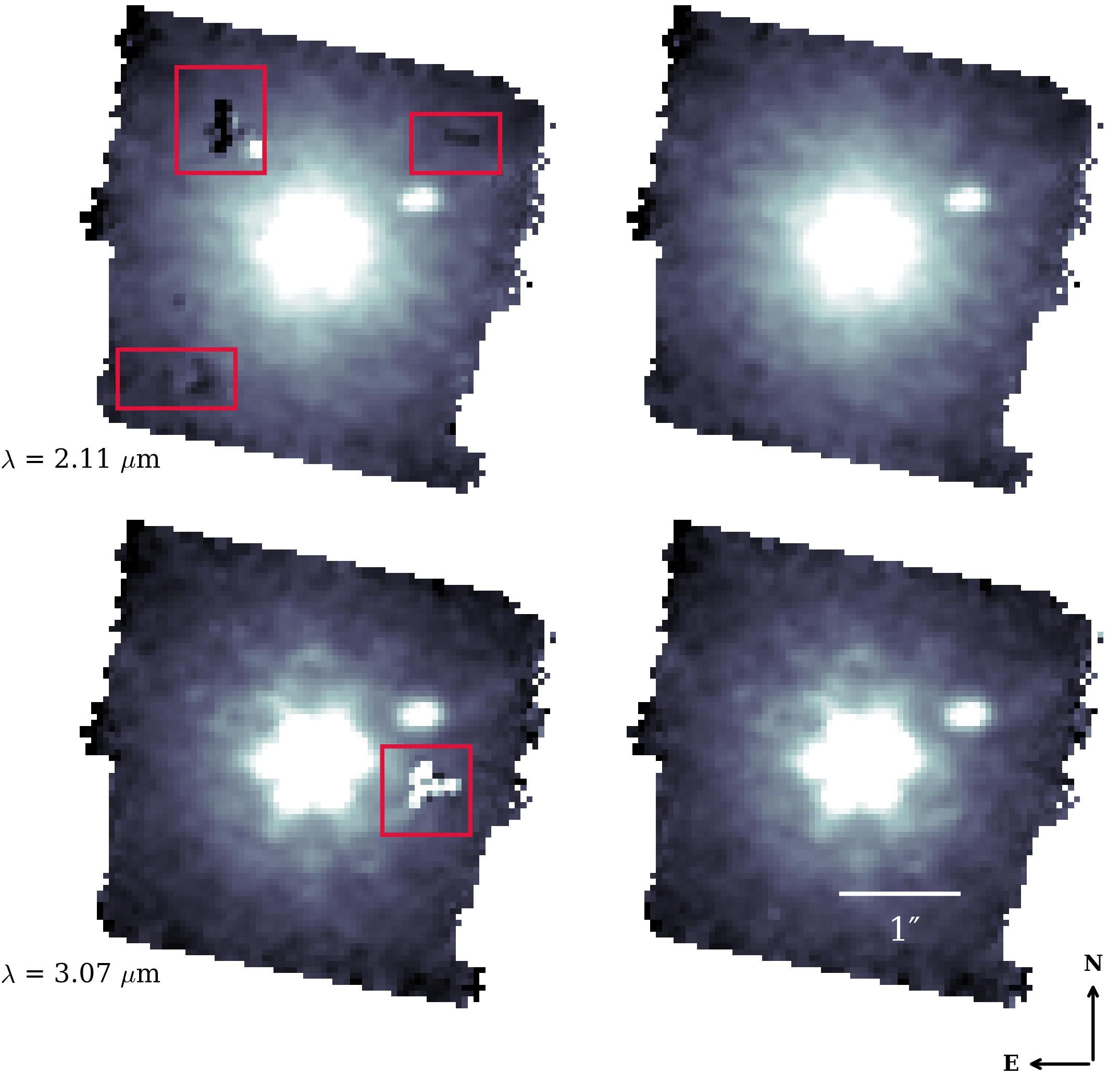}
    \caption{Images of M87 taken at two slices of the data cube when no additional bad pixels are flagged in the 2D detector images before building the cube (left) and the same two slices of the cube after applying our supplemental masking of bad pixels in each of the calibrated detector images prior to assembling the data cube (right). The red boxes in the left images denote positive and negative artifacts that remain the final cube, but that do not appear in the data cube on the right when using our approach. The wavelength of each slice is given at the bottom of the left images, the spaxels are 0\farcs05 in side, and the images are oriented with north up. The feature to the northwest of the M87 nucleus is a jet knot.}
  \label{fig:cleancube}
\end{singlespace}
\end{figure}

As an example, in the left set of images in Figure \ref{fig:cleancube}, we show two slices of the final M87 data cube when the DQ array is not adjusted and additional bad pixels (beyond what is caught by the JWST pipeline, even with snowball detection turned on) are not flagged. There are many extended, irregularly shaped negative and positive artifacts in the data cube and this occurs across all wavelengths. In contrast, the right set of images in Figure \ref{fig:cleancube} displays the same two slices of the final data cube when we mask remaining bad pixels in the 2D detector image for each exposure before building the combined cube. For this figure, all parts of the data reduction are the same with the exception of whether supplemental bad pixels that we identified are masked in the 2D detector images before generating the data cube. As can be seen, there is a substantial improvement in our final data cube.

\subsection{Building a Data Cube} \label{subsec:cube_building}

We further make an adjustment in how the exposures are merged when building the combined data cube to mitigate artifacts near the detector gap. Other analyses (e.g., \citealt{Wylezalek_2022, Ishikawa_2025}) have also elected to modify the standard procedure for producing a merged data cube, and in some cases \citep{Ruffio_2024} a data cube is not even created. In our case, we generate our final data cube by first using the pipeline's cube-building step, inputting the calibrated detector images from the 16 dithered exposures, requiring the spaxels to be 0\farcs05 on a side, and keeping the other default settings. This allows us to determine the dimensions of an output cube and the RA and Dec of the center. 

We then construct individual cubes for each exposure, again using the pipeline's cube-building step, but we require each exposure be drizzled onto the common 3D grid that was previously determined. We also make a minor alteration to the pipeline's cube-building code so that the drizzle weights of the voxels in the individual cubes are stored. We combine the 16 individual cubes together, and perform additional masking, requiring all voxels to have a DQ flag of `\texttt{GOOD}', a positive drizzle weight, and an intensity that is finite, positive, and nonzero. Non-satisfying voxels are masked during the merger by designating their intensity to be NaN and their drizzle weight to be $0$. Furthermore, for each spaxel in each individual cube, we identify the apparent boundaries of the gap between the NRS1 and NRS2 detectors by locating the longest contiguous region of spectral pixels with NaN values and $0$ drizzle weights. The union of these regions across the 16 dithered exposures was determined for each spaxel and masked when building the final cube, again by designating the intensity to be NaN and the summed drizzled weight to be $0$. We note that because the individual cubes have a common grid and the drizzle weights from the pipeline's cube-building step are stored, no additional interpolation step is needed when the individual cubes are merged together.

\begin{figure}[h!]
\begin{singlespace}
  \centering
  \includegraphics[width=\linewidth]{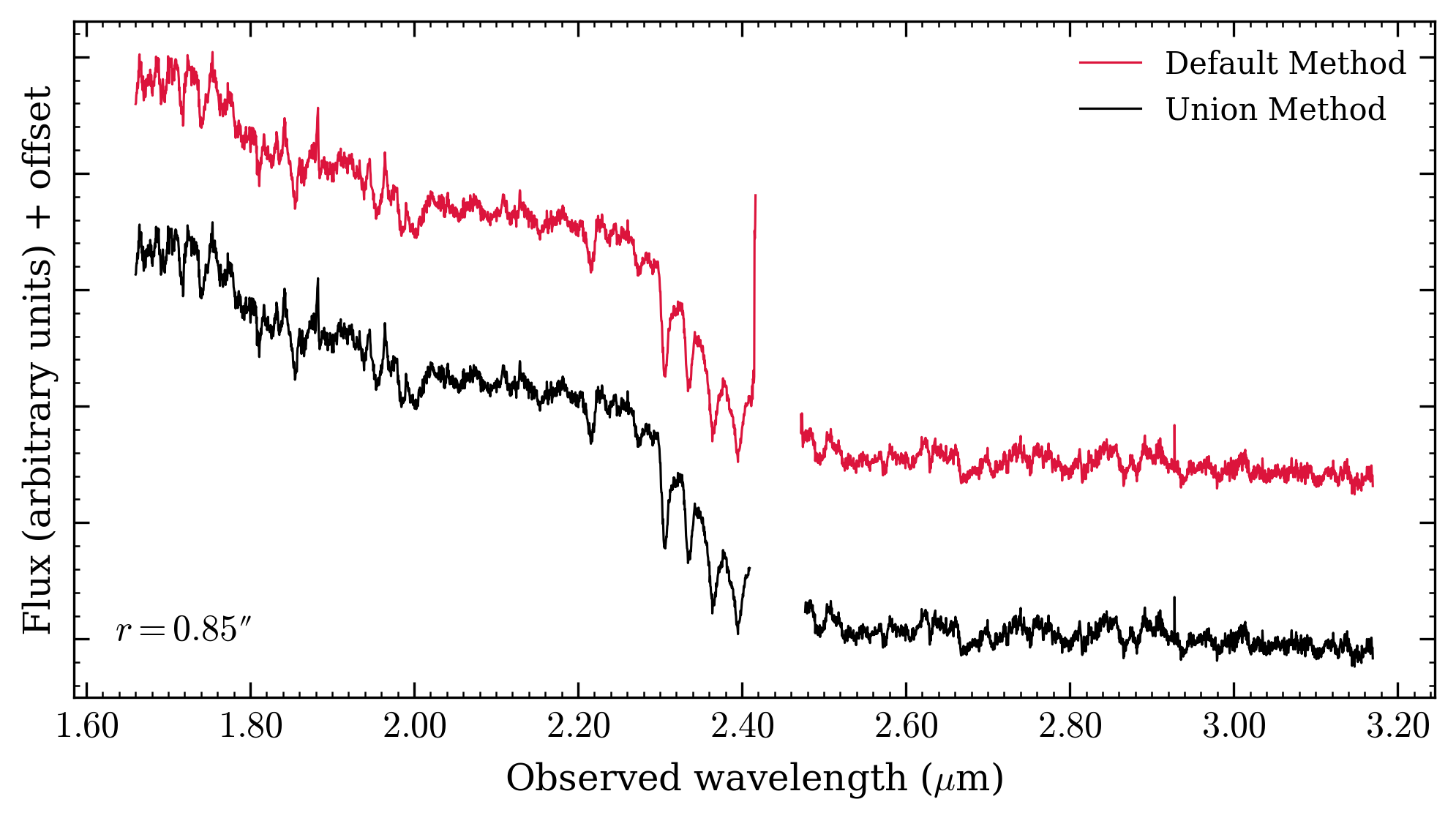}
    \caption{A comparison of the spectrum extracted from a single 0\farcs05 spaxel in the final M87 data cube when using the standard method (red) and our union method (black) for building the cube. With the typical approach, artifacts are commonly seen adjacent to the NRS1/NRS2 detector gap. The detector gap is the white space with missing data, at $\sim$$2.45\ \mu$m above. With the union method, the spectra are clean and enable the secure measurement of the stellar LOSVD from the CO bandheads near the detector gap.}
  \label{fig:detectorgap_artifacts}
\end{singlespace}
\end{figure}

With our union method for assembling the data cube, the final cube has minimal artifacts in the spectral region near the detector gap and many fewer spaxels, particularly at the spatial edges of the cube, contain large amounts of NaN values. Eliminating artifacts by the detector gap is especially important, as in Section \ref{sec:stellar_kin} we measure the detailed shape of the stellar LOSVD from the CO bandheads in this spectral region. Figure \ref{fig:detectorgap_artifacts} demonstrates the spurious features that can appear around the detector gap when we use the typical procedure. For comparison, we plot the spectrum from the same spaxel in our final cube when adopting our union method. For this figure, all aspects of the data reduction process are identical, including the additional masking of bad pixels in 2D detector images, and the only difference is in how the merged data cube is constructed.

While one could spectrally mask anomalies near the detector gap as a post-processing step when measuring the stellar kinematics, we find that the artifacts are unpredictable, occurring at varying spectral locations away from the detector gap and impacting the CO bandheads differently across the NIRSpec field-of-view (FoV). Our tests using the default method for assembling the final data cube and applying a spectral mask to exclude the artifacts near the detector gap when fitting the spectra revealed that the width of the spectral mask had an impact on the inferred stellar LOSVDs for M87. Instead, our union method for building the final cube yields clean spectra, removes the need for spectral masking near the detector gap, and results in robust stellar kinematic measurements.

\begin{figure*}
\begin{singlespace}
  \centering
  \includegraphics[width=\linewidth]{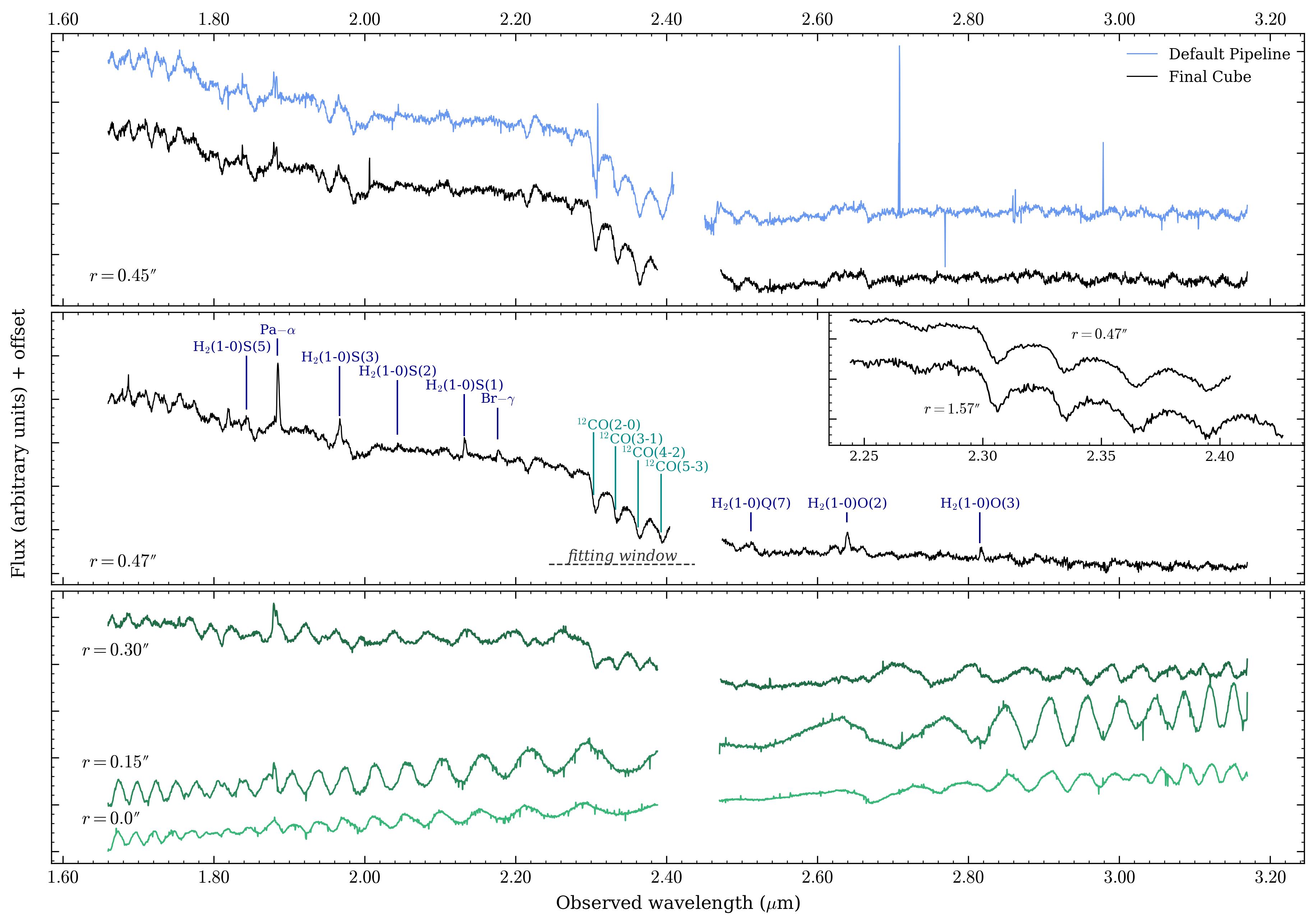}
\caption{\emph{Top:} A spectrum from a single 0\farcs05 spaxel in the M87 data cube constructed using the default JWST pipeline (light blue) and in the final data cube generated using the modifications to the pipeline described in Sections \ref{sec:obs_reduction}, \ref{subsec:removing_outliers}, and \ref{subsec:cube_building} (black). The latter spectrum exhibits noticeable improvements relative to the former. \emph{Middle:} An example spectrum from our final data cube extracted from a single 0\farcs05 spaxel just outside the nucleus, with the emission lines labeled in dark blue and the $^{12}$CO bandhead absorption features labeled in teal. The horizontal dashed gray line indicates the wavelength range over which the CO bandheads were fit, as described in Section \ref{sec:stellar_kin}. A zoom-in of this spectral region for the same spectrum is shown in the top right inset of the panel, along with a second spectrum extracted from a single spaxel at near the edge of the data cube. Even in single spaxels, just outside the nucleus and at the edge of the FoV, the $S/N$ is very high. \emph{Bottom:} Typical nuclear spectra from single spaxels in the final M87 data cube. Wiggles are clearly seen, despite our 16 exposure, small-cycling dither pattern. The wiggles improve with increasing distance from the center, and are no longer observed in the top two panels, at $r$$\sim$0\farcs45.}
  \label{fig:summary_cube}
\end{singlespace}
\end{figure*}

\section{The M87 Data Cube} \label{sec:m87cube}

After completing the data reduction, including implementing all of the changes to the default pipeline described in Sections \ref{sec:obs_reduction}, \ref{subsec:removing_outliers}, and \ref{subsec:cube_building}, our final cube covers the central $3\arcsec\times 3\arcsec$ region of M87 with high angular resolution and a sampling of $0\farcs05$ pixel$^{-1}$ along the spatial $x$ and $y$ axes. The wavelength axis spans $1.66-3.17\ \mu$m with a scale of $3.96\times10^{-4}\ \mu$m pixel$^{-1}$ and a spectral resolving power of $R\sim2700$. In the top panel of Figure \ref{fig:summary_cube}, we compare a spectrum extracted from a single $0\farcs05$ spaxel from the cube generated with the default JWST pipeline and the same spaxel from our final data cube. While Figures \ref{fig:cleancube} and \ref{fig:detectorgap_artifacts} focused on isolating a specific modification we made to the standard reduction, the top panel of Figure \ref{fig:summary_cube} compares the outcome when the default JWST pipeline is run from start to finish versus when all of our changes are implemented together.

The NIRSpec data have spectacular $S/N$ even in individual spaxels near the nucleus and at the edge of the data cube, with values typically $\sim$$150$ (see Section \ref{sec:stellar_kin}). The $K$-band CO bandhead absorption features are prominent throughout FoV and given M87's redshift we find between two and four of the $^{12}$CO bandheads blueward of the detector gap, as the spectral location of the gap varies across the FoV. These $K$-band CO bandheads serve as the primary tracer for measuring the stellar kinematics. We also find numerous emission lines, which previously have gone undetected in the near-infrared (e.g.,~\citealt{Gebhardt2011}), highlighting the sensitivity of JWST and the benefits of space-based observations. The emission features we detect include multiple warm molecular hydrogen (H$_2$) lines\footnote{https://www.gemini.edu/observing/resources/near-ir-resources/spectroscopy/important-h2-lines}, as well as a few hydrogen recombination lines\footnote{https://www.gemini.edu/observing/resources/near-ir-resources/spectroscopy/hydrogen-recombination-lines}, the strongest of which is Pa$\alpha$. The middle panel of Figure \ref{fig:summary_cube} presents spectra from single spaxels and illustrates the high $S/N$ and the absorption and emission lines present. In addition to the emission lines labeled in Figure \ref{fig:summary_cube}, at other spatial locations we also detect H$_2$($3-2$)S($7$), H$_2$($1-0$)S($0$), H$_2$($1-0$)Q($5$), and H$_2$($1-0$)O($4$). An analysis of the gas kinematics will be the subject of future work.

\subsection{Wiggles in the Nuclear Spectra} \label{subsec:wiggles}

Although the M87 data have high $S/N$ and strong absorption and emission lines, the spectra at the nucleus show unphysical periodic amplitude modulations. These ``wiggles'' appear as a result of undersampling of the PSF and are discussed at length by \cite{Law2023}. While \cite{Law2023} demonstrate that a four-point dither pattern reduces the undersampling artifacts, we see that even with our 16 exposure, small-cycling dither pattern, the wiggles remain. Likewise, \cite{Ruffio_2024} find that a nine-point dither pattern did not result in as significant of an improvement in the wiggles as desired, and they further discuss how the effectiveness of the dither pattern is field- and wavelength-dependent. Beyond increasing the number of dithers, perhaps the preset dither pattern itself can be re-examined and refined. In some cases, the exponential modified-Shepard method (EMSM) is used when generating the data cube in order to reduce the wiggles \citep{Marshall_2023, Perna_2023}. However, EMSM data cubes have slightly lower spatial and spectral resolution relative to the drizzled versions \citep{Law2023}, and thus we continue to use drizzle weighting.

In the bottom panel of Figure \ref{fig:summary_cube}, we show representative spectra from single spaxels at increasing projected radii, $r$, from the nucleus. The wiggles become less problematic with increasing $r$, and no longer are an issue by $r \sim 0\farcs45$. While several packages have been released, outside of the JWST pipeline, to correct the wiggles \citep{Perna_2023, Dumont_2025, Shajib_2025} and we have experimented with our own procedures, we find the extracted stellar kinematics for M87 to be sensitive to the exact methods adopted. Moreover, the impact of the AGN can be seen in the innermost spectra, specifically in the dilution of the CO bandheads and the shape of the continuum, but the AGN is no longer an issue by $r \sim 0\farcs45$. For these reasons, we choose to focus on the spectra at $r \geq 0\farcs45$ in this paper. Investigation of the innermost region of the data will be the subject of future work.

\section{Spatial Masking and Binning} \label{sec:masking_binning}

Given the wiggles and the impact of the AGN on the spectra at the nucleus, we apply a circular mask with $r<$ 0\farcs45 that is centered on the spaxel with the largest intensity when summing over the wavelength axis of the data cube. We mask the jet knot to the northwest of the nucleus based on visual inspection of the collapsed data cube and the presence of possible wiggles in the spectra at the location of the jet knot. The outermost edges of the data cube were also masked such that there are no NaN values for the intensity within the wavelength range over which we fit the stellar kinematics (see Section \ref{sec:stellar_kin}) and there is a single contiguous field of spaxels from which we measure the stellar LOSVDs.

We use the Voronoi binning method \cite{Cappellari2003} to construct spatial bins and ensure all the spectra have the necessary high $S/N$ to reliably measure the stellar kinematics. Because we have very high $S/N$ in single spaxels over much of the FoV, when applying the Voronoi binning method, only the outer regions are binned and the bin sizes remain fairly small. After a round of Voronoi binning, we adjust the edges of the spatial mask to exclude bins that still did not meet our bin $S/N$ threshold and smooth the irregularly shaped edges of the updated mask. The Voronoi binning is repeated, ultimately resulting in 2810 bins, of which 2232 are composed of a single spaxel.

\begin{figure*}
\begin{singlespace}
  \centering
  \includegraphics[width=\linewidth]{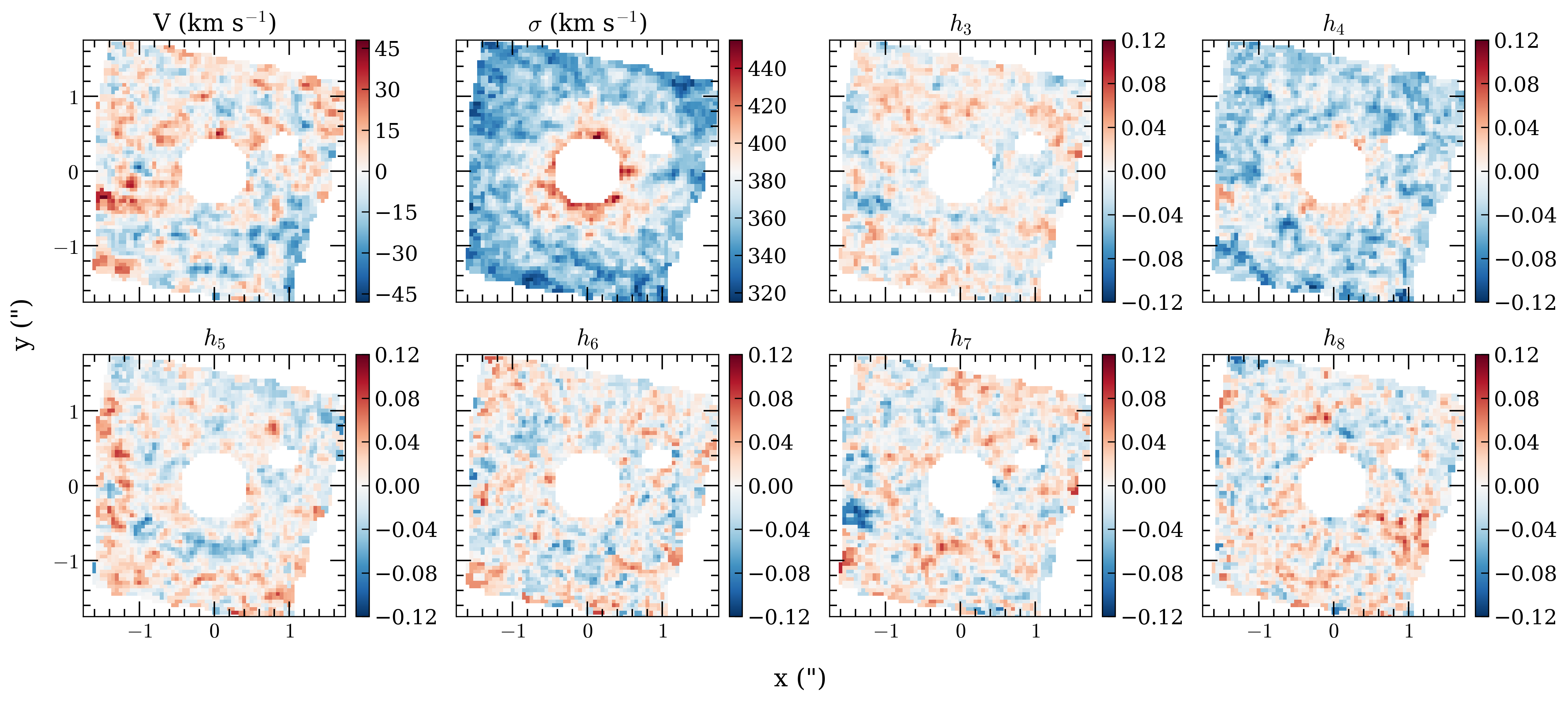}
    \caption{The first eight GH moments used to describe the LOSVD are shown across the NIRSpec FoV. There are 2180 spatial bins for which a kinematic measurement was made, and 2232 of the spatial bins are composed of a single 0\farcs05 spaxel. The central 0\farcs45 and the jet knot to the northest have been masked. The maps are oriented with north up and east to the left.}
  \label{fig:kinmaps}
\end{singlespace}
\end{figure*}

\section{Stellar Kinematics} \label{sec:stellar_kin}

We use the penalized pixel fitting method (pPXF; \citealt{Cappellari2023}) to compare velocity template stars from the PHOENIX synthetic library \citep{Husser2013}, convolved with an LOSVD and adjusted by degree 3 multiplicative Legendre polynomial, to the observed galaxy spectrum in each spatial bin. The LOSVD is parameterized in terms of a Gauss-Hermite (GH) series, and we report the radial velocity $V$, velocity dispersion $\sigma$, and higher-order GH moments ($h_3-h_8$), which quantify the asymmetric and symmetric deviations of the LOSVD from Gaussian. We fit over a wavelength range focused on the CO bandheads located on the NRS1 detector and spectrally mask the \ion{Ca}{1} absorption that is not well matched by our PHOENIX template library. The details of the template library, the masking of \ion{Ca}{1}, the wavelength range we fit over, and other spectral fitting parameters are discussed in Section \ref{subsec:kintests} below.

For every spatial bin, after an initial fit to the observed M87 spectrum with a pPXF bias parameter of 0.2, we determine the uncertainties on the stellar kinematics using a Monte Carlo procedure. During each of the 200 Monte Carlo realizations, we add random Gaussian noise to the initial best-fit pPXF model, such that the noise is equal to the standard deviation of the model residuals. We re-fit with pPXF with the bias turned off and record the stellar kinematics from each realization. With the 200 iterations complete, we calculate the median and standard deviation of the resulting distribution for each GH moment, which we take to be the kinematic values that we report throughout the paper and their 1$\sigma$ uncertainties.

Figure \ref{fig:kinmaps} shows the maps of the first eight GH moments. We do not symmetrize the maps to remove overall offsets in the odd GH moments and average away outliers, as is common practice prior to dynamical modeling (e.g.,~\cite{Walsh_2015, Roberts_2021, Thater_2022}). However, we do subtract offsets in the odd GH moments, and the offsets were determined by calculating the median value over all spatial bins. For $V$, this offset corresponds to the galaxy's systemic velocity and for the higher odd moments the offsets are typically attributed to template mismatch. We find very small offsets for $h_3$, $h_5$, and $h_7$, of $-0.002$ to $-0.006$, indicating minimal template mismatch. 

\begin{figure*}
\begin{singlespace}
  \centering
  \includegraphics[width=\linewidth]{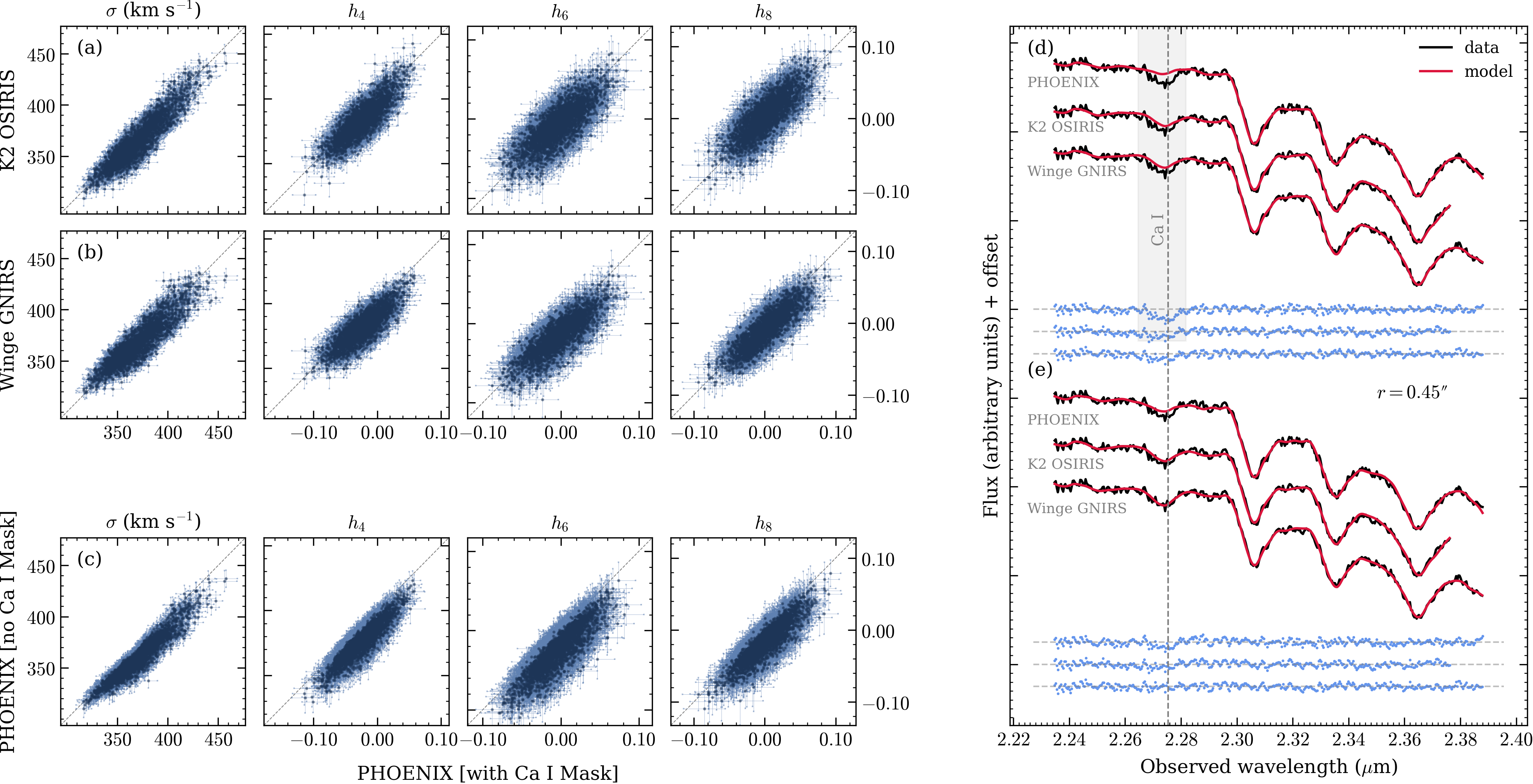}
    \caption{\emph{Left:} Panels (a) and (b) compare even GH moments derived using the K2 OSIRIS and Winge GNIRS libraries to those from the PHOENIX library, all with \ion{Ca}{1} masked. Panel (c) contrasts moments obtained with and without masking \ion{Ca}{1}, using PHOENIX templates. The dotted line indicates one-to-one correspondence. When \ion{Ca}{1} is masked, the GH moments are consistent across all libraries. However, omitting the mask with PHOENIX introduces a systematic offset in the kinematics. Although not shown, when using K2 OSIRIS and Winge GNIRS consistent kinematics are found regardless of whether \ion{Ca}{1} is masked. \emph{Right:} Panels (d) and (e) show spectral fits to the same M87 spectrum (black) using PHOENIX, K2 OSIRIS, and Winge GNIRS templates, with \ion{Ca}{1} masked (gray band) and unmasked, respectively. The red lines show best-fit pPXF models, with residuals in blue. All models are a good match to data when \ion{Ca}{1} is masked. When \ion{Ca}{1} is not masked, the PHOENIX templates cannot reproduce \ion{Ca}{1} well, whereas the two empirical libraries are able to do so. While the same single spaxel M87 spectrum is shown, the ending wavelengths differ due to the varying spectral coverage of the template libraries, with K2 OSIRIS cutting off earlier than the others.}
  \label{fig:templib}
\end{singlespace}
\end{figure*}

\begin{figure*}
\begin{singlespace}
  \centering
  \includegraphics[width=\linewidth]{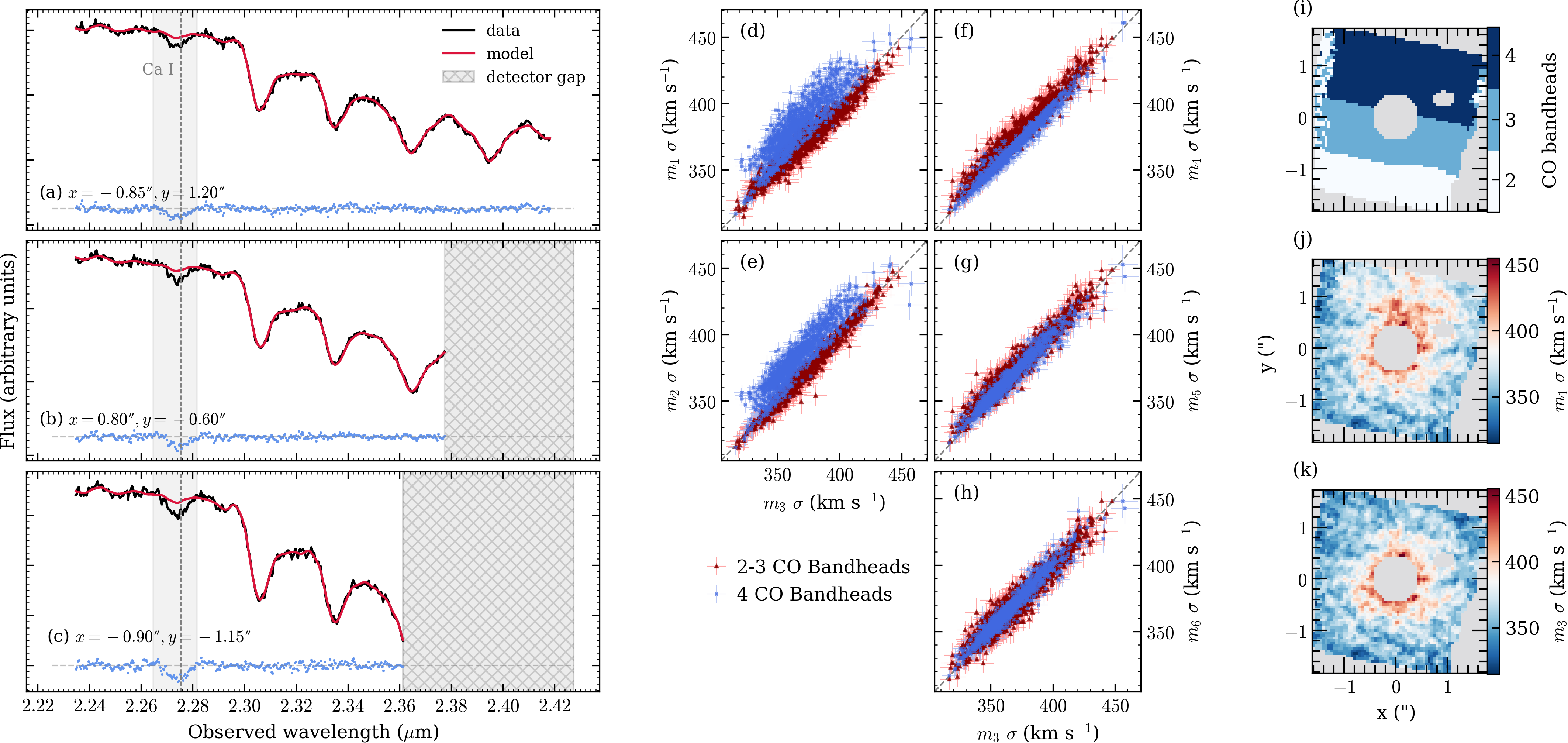}
    \caption{\emph{Left:} Panels (a)$-$(c) present example M87 spectra (black) from single spaxels at different spatial locations, best-fit pPXF models (red), and residuals (blue). The \ion{Ca}{1} feature was masked (gray band). There are two to four CO bandheads within the fitting region due to the spectral variation of the detector gap (hatched band). \emph{Middle:} Panels (d)$–$(h) compare velocity dispersions measured when using polynomial degrees $1–2$ and $4–6$ relative to degree $3$. Measurements made from spectra with two or three CO bandheads (red points) follow the one-to-one line (dashed black line) in panels (d)$-$(h), but measurements from spectra with four CO bandheads (blue points) only fall on the one-to-one line in panels (f)$-$(g). At least a degree $3$ polynomial is needed for accurate fits over the broader wavelength range with four CO bandheads, whereas results are consistent across degrees when fitting the narrower wavelength range with two or three bandheads. \emph{Right:} Panel (i) maps the number of CO bandheads on the NRS1 detector and panels (j) and (k) show velocity dispersions when using degree $1$ and degree $3$ polynomials, respectively. Since the number of CO bandheads varies systematically across the FoV and using too low of a polynomial order causes the dispersion to be overestimated when four CO bandheads are fit, there is an artificial gradient in the dispersion map in panel (j), but using a degree $3$ polynomial resolves this issue. The same behavior was seen for the other even GH moments.}
  \label{fig:mdegreeplot}
\end{singlespace}
\end{figure*}

From the maps, we find rotation at the center of M87, with $V \sim \pm 45$ \kms, the velocity dispersion sharply rises towards the center reaching $\sim$$420$ \kms\ at $r = $ 0\farcs45, a slight elevation in $h_4$ toward the nucleus, and the other higher GH moments fluctuate about zero. Representative uncertainties are $5$ \kms\ and $7$ \kms\ for $V$ and $\sigma$, respectively, and $0.01-0.02$ for $h_3-h_8$. The $S/N$, taken to be the median of the M87 spectrum over the common wavelength range fit among all the spatial bins (of $2.2443-2.3632\ \mu$m) divided by the standard deviation of the pPXF residuals, spans $S/N=116-223$ with a median value of $166$. Example spectral fits at different spatial locations can be found in panel (d) of Figure \ref{fig:templib} (the top spectrum) and in panels (a)$-$(c) in Figure \ref{fig:mdegreeplot}.

\subsection{Robustness of the Stellar Kinematics} \label{subsec:kintests}

We examine the robustness of the fiducial stellar kinematics and provide details about these tests below.

\subsubsection{Template Libraries and Spectral Masking of Ca I} \label{subsubsec:veltemps}

For the fiducial kinematic measurement, we select 90 stars from the high-resolution PHOENIX library \citep{Husser2013}, with effective temperatures of $\mathrm{T_{eff}}=2400-6500$ K, surface gravities of $\log(g)=0-5$, metallicities of $-0.5\leq[\mathrm{Fe/H}]\leq +0.5$ dex, and solar alpha element abundances. The chosen properties are representative of G, K, and M giant, supergiant, and dwarf stars with a range in metallicity. The PHOENIX library is commonly used to analyze kinematics in nearby galaxies observed with JWST \citep{Tahmasebzadeh_2025, Nguyen_2025, Dumont_2025, Taylor_2025} because it spans a wide wavelength range at high spectral resolution and many stars with a variety of properties can be sampled. However, the synthetic spectra assume a particular stellar atmospheric code. Although empirical template libraries are more limited in wavelength coverage, $S/N$, and number of stars, comparing results from synthetic and empirical libraries is essential, especially because using empirical templates has been the standard practice for pre-JWST kinematic measurements.

In addition, when using the PHOENIX library, we observe that the \ion{Ca}{1} absorption in M87 is not matched well, and thus we mask the line. \cite{Krajnovic_2009} note that properly fitting \ion{Ca}{1} requires cool dwarf stars. While our PHOENIX library does include such stars, we still do not find a good match to the feature. Other work using empirical template libraries that lacked cool dwarf stars also had difficulty fitting \ion{Ca}{1} and chose to mask the line (e.g., \citealt{Walsh_2016, Walsh_2017}). 

Given that PHOENIX is a synthetic library and the M87 \ion{Ca}{1} feature is not fit well, we explored other empirical template libraries, including a set of stars we observed with the OH-Suppressing Infrared Integral Field Spectrograph (OSIRIS) aided by AO when the instrument was on the Keck II telescope, which we will refer to as the K2 OSIRIS library, and the \cite{Winge2009} Gemini Near-Infrared Spectrograph (GNIRS) library. The K2 OSIRIS library \citep{Walsh_2012} is composed of 22 K and M giant stars and G and K dwarf stars observed with the broadband $K$ filter over $1.97-2.38\ \mu$m with $R \sim 3800$. The Winge GNIRS library contains 23 G, K, and M giant, supergiant, and dwarf stars observed with GNIRS in IFU mode over both a blue and red spectral setup, covering $2.15-2.43\ \mu$m (in air) at $R \sim 7600$. As part of the data reduction, \cite{Winge2009} fit the continuum shape of each template star and removed it. Thus, we find a slightly larger multiplicative polynomial degree of 4 is needed when using the Winge GNIRS library to match the M87 NIRSpec data compared to when using the PHOENIX and K2 OSIRIS libraries, both of which have template spectra with the continuum shape intact.

When fitting the M87 NIRSpec data in the same manner as described in Section \ref{sec:stellar_kin}, including masking \ion{Ca}{1}, but adopting the K2 OSIRIS template library, we recover consistent kinematics as shown in panel (a) of Figure \ref{fig:templib}. Likewise, using the Winge GNIRS library, continuing to mask \ion{Ca}{1} in the M87 data, and using a multiplicative polynomial of degree 4, yield comparable stellar kinematics as the fiducial run, as can be seen in panel (b) of Figure \ref{fig:templib}. While Figure \ref{fig:templib} displays only the even moments, we see similar results for the odd moments with comparable scatter about the one-to-one line. Example spectral fits to the same M87 spectrum, but using the three template libraries, are given in Figure \ref{fig:templib} panel (d). The best-fit pPXF model reproduces the observed M87 spectrum nicely over the wavelengths fit. The masked \ion{Ca}{1} line is not matched when using the PHOENIX library [top spectrum of panel (d)] and is better recovered when using the K2 OSIRIS and Winge GNIRS libraries [middle and bottom spectra of panel (d)] even though the data in this spectral region were not part of the fit.

Next, we repeat the measurement with the three template libraries, but did not mask \ion{Ca}{1}. With both the K2 OSIRIS and Winge GNIRS libraries, the stellar kinematics are identical regardless of whether \ion{Ca}{1} is masked or included in the fit, however there is a clear offset in the even GH moments when the PHOENIX library is used and the \ion{Ca}{1} is masked or not masked [Figure \ref{fig:templib} panel (c)] -- the even GH moments are systematically biased low when \ion{Ca}{1} is not masked. More specifically, the distribution of velocity dispersion differences ($\sigma$ extracted when \ion{Ca}{1} is not masked $-$ $\sigma$ recovered with the fiducial settings) over all spatial bins is systematically shifted and skewed, ranging from $-17$ \kms\ to $-5$ \kms\ (the 68\% interval) with a median difference of $-9$ \kms. The differences become worse at high dispersions and span from $-25$ \kms\ to $-6$ \kms\ (68\% interval) with a median of $-16$ \kms\ for bins with a fiducial $\sigma > 400$ \kms. Both systematic shifts are outside the typical statistical uncertainty of $\sim7$ \kms\ derived from the Monte Carlo simulation for $\sigma$ in Section \ref{sec:stellar_kin}. Figure \ref{fig:templib} panel (e) illustrates the same M87 spectrum as in panel (d) and the best-fit pPXF model with the PHOENIX, K2 OSIRIS, and Winge GNIRS libraries when \ion{Ca}{1} is not masked. Again, \ion{Ca}{1} is poorly fit by the PHOENIX library, while the two empirical libraries are capable of matching \ion{Ca}{1}, and can do so even better than in panel (d), now that this spectral region is part of the fit.

Since both empirical libraries reproduce the \ion{Ca}{1} feature, and yield consistent kinematics whether \ion{Ca}{1} is masked or not, coupled with the stability in the kinematics regardless of the three template libraries (PHOENIX, K2 OSIRIS, Winge GNIRS) adopted, as long as \ion{Ca}{1} is masked, we conclude that using the PHOENIX library and masking \ion{Ca}{1} produces robust stellar kinematics.

\subsubsection{Additive and Multiplicative Polynomials} \label{subsubsec:add_mult_deg}

Often additive polynomials are used to aid in modeling an AGN component or account for sky subtraction errors, whereas multiplicative polynomials are employed to correct for small errors in spectral flux calibration \citep{Cappellari2023}. In Section \ref{sec:stellar_kin}, we adopt a multiplicative Legendre polynomial of degree $3$ to modify the shape of the LOSVD-convolved template stars. Although M87 harbors a low-luminosity AGN, our spectral fitting is restricted to the region outside of the point source, at $r \geq$ 0\farcs45, and we should not need an additive polynomial. For completeness, we conduct tests with an additive polynomial of degree $0$ and $1$, but these reveal elevated values for $h_6$ and $h_8$ of $\sim0.10-0.15$ over the entire FoV that are not realistic. Hence, we continue without an additive polynomial.

When varying the multiplicative polynomial degree between $1-6$, we recover consistent kinematics when there were two and three CO bandheads before the detector gap, but find we require a multiplicative polynomial with degree $\geq 3$ when there were four bandheads before the detector gap. This behavior and its implications can be seen in Figure \ref{fig:mdegreeplot}. As shown in panels (a)$-$(c) of Figure \ref{fig:mdegreeplot}, since we use the same starting wavelength but differing ending wavelengths for the fit that go to the detector gap in each spatial bin, there is anywhere from two to four CO bandheads being fit and there is varying wavelength ranges over which the fit is performed.

Panels (d)$-$(h) in Figure \ref{fig:mdegreeplot} compare the dispersion measured when using a degree $1-2$ [panels (d) and (e)] and $4-6$ [panels (f)$-$(g)] to the value determined when using a degree $3$ polynomial. The red points correspond to spectra with two and three CO bandheads, and in all panels, (d)$-$(h), the red points fall along the one-to-one line. In other words, $\sigma$ is stable and multiplicative degree $1-2$ and $4-6$ polynomials return similar dispersions as when a degree $3$ polynomial is used. In contrast, the blue points correspond to spectra with four CO bandheads and these dispersions in panels (d) and (e) are biased high, above the one-to-one line, but lie along the one-to-one line in panels (f)$-$(h). Therefore, when four bandheads, and a wider wavelength range, are fit using too low of a multiplicative polynomial degree, we measure an inflated $\sigma$. Once a large enough polynomial degree is used, the dispersions are consistent and there is no dependence on the multiplicative polynomial degree.

Because the spectral location of the detector gap varies across the NIRSpec FoV, there is a spatial dependence on the number of CO bandheads being fit, as displayed in panel (i) of Figure \ref{fig:mdegreeplot}. The southern portion of IFU has spectra with two CO bandheads fit and the northern part of the IFU has spectra with four CO bandheads fit. Consequently, when too low of a multiplicative polynomial degree is adopted, as in panel (j) of Figure \ref{fig:mdegreeplot}, there is a false gradient detected in $\sigma$, with enlarged dispersions in the northern part of the IFU where there are four CO bandheads present. With a degree 3 multiplicative polynomial, the spatial gradient in $\sigma$ is eliminated [panel (k)] and the spatial trend in the dispersion, with the dispersion rising toward the nucleus, is real. While Figure \ref{fig:mdegreeplot} focuses on the velocity dispersion, we found the same is true for the other even GH moments, $h_4$, $h_6$, and $h_8$.

\subsubsection{Spectral Fitting Window} 

As there is only one or two very weak CO bandheads seen at the start of the NRS2 detector, when running pPXF we fit over a wavelength range starting at $2.2347\ \mu$m and fit up to the detector gap. Such a fitting window has the benefit of targeting the NRS1 CO bandheads along with sufficient continuum, being free from prominent emission lines, and multiple velocity template libraries cover this spectral region. A similarly focused wavelength range was fit for other nearby galaxies observed with NIRSpec \citep{Tahmasebzadeh_2025, Nguyen_2025, Taylor_2025}.

We adjusted the starting wavelength, testing values of $2.2458$, $2.2605$, and $2.2855\ \mu$m, and continued to fit to the detector gap for each spatial bin. In general, we find consistent results as the fiducial kinematics presented in Section \ref{sec:stellar_kin}, but there is a systematic deviation in the even moments when using a starting wavelength of $2.2855\ \mu$m relative to the fiducial kinematics in spatial bins with high velocity dispersions. For example, in bins with a fiducial $\sigma > 400$ \kms, the difference ($\sigma$ found with a fitting wavelength starting at $2.2855\ \mu$m $-$ $\sigma$ determined with the fiducial settings) ranges from $-22$ \kms\ to $1$ \kms\ (the 68\% interval) with a median of $-11$ \kms. In the case of using $2.2855\ \mu$m for the starting wavelength, the kinematics appear to be less stable, likely due to the limited and less well-defined continuum before the start of the very broad $^{12}$CO(2$-$1) bandheads in M87. Exploring wider wavelength ranges, such as fitting over the NRS1 detector or the entire NRS1$+$NRS2 wavelength range requires a thorough examination of the most appropriate polynomial to use (see Section \ref{subsubsec:add_mult_deg}) and the spectral masking of emission lines. We do not include such a fit here.

\subsubsection{Number of GH Moments} 

Many prior kinematic studies that describe the LOSVD with a GH series use four, or occasionally six, moments (e.g.,~\citealt{Merrell_2023, Thater_2023}), however given the high $S/N$ of the NIRSpec data we report the first eight GH moments. Recent studies with high $S/N$ spectra also utilize eight GH moments \citep{Pilawa2022, Liepold2023, Liepold_2025, Pilawa_2025}, and \cite{Mehrgan_2023} emphasize that nonparametrically-derived LOSVDs observed for massive early-type galaxies require at least six or eight GH moments to capture the detailed structure of the distributions. \cite{Liepold_2020} demonstrated for the massive early-type galaxy NGC 1453 that truncating the GH series early can result in inflated values for the last even moment measured, and that continuing the series out to higher terms resolves the issue and the additional higher even moments then scatter about $0$. 

We explore fitting $4-14$ GH moments and find consistent results as the fiducial kinematics presented in Section \ref{sec:stellar_kin} for the moments in common between the runs. Therefore, here we do not encounter a situation where the last even moment is elevated if the GH series is truncated at a low order. When fitting two GH moments and comparing to the fiducial kinematics, we find a slight systematic shift mainly at high dispersions. In spatial bins with a fiducial $\sigma > 400$ \kms, the difference ($\sigma$ when fitting two GH moments $-$ $\sigma$ from the fiducial settings) ranges from $-15$ \kms\ to $0$ \kms\ (the 68\% interval) with a median of $-6$ \kms. We continue to present the the first eight GH moments because the high-quality NIRSpec data enable such a measurement and the M87 large-scale stellar kinematics from Keck KCWI were extracted in the same way \citep{Liepold2023}; we will fit orbit-based, triaxial dynamical models to both the NIRSpec small-scale and the KCWI wide-field stellar kinematics in the future.

\subsubsection{Relative Weights of the Template Stars} 

To derive our fiducial kinematics, we fit pPXF to each spatial bin separately, allowing the weights applied to the template stars to change between bins. Another approach is to require the relative weights of the template stars to remain fixed between spatial bins when measuring the kinematics. We carry out this second method and determine the relative template weights through an initial fit to a very high $S/N$ spectrum of M87, constructed by summing together all the M87 spectra within a central, circular annulus with an inner and outer radius of 0\farcs45 and 0\farcs7. The resulting optimal template was then used as the reference when comparing to the observed M87 spectrum in each spatial bin. We find kinematics that are consistent with our fiducial run.

\subsubsection{Spectral Resolution} 

We assume the NIRSpec data have a constant $R$ of $2700$, or an instrumental dispersion of $\sigma_\mathrm{inst} = 47$ \kms, following the JWST User Documentation\footnote{https://jwst-docs.stsci.edu/jwst-near-infrared-spectrograph/nirspec-instrumentation/nirspec-dispersers-and-filters} (JDox) for the G235H grating at our data cube's central wavelength of $2.415\ \mu$m. Recently, \cite{Shajib_2025b} found that the in-flight spectral resolution exceeded the JDox estimates by $5-45\%$ across all NIRSpec configurations and wavelengths. If instead we adopt $R = 3289$, calculated using \cite{Shajib_2025b} for the F170LP/G235H IFU configuration at a wavelength of $2.415\ \mu$m, we obtain identical kinematics as the fiducial set in Section \ref{sec:stellar_kin}.

\section{Multiscale Kinematic Comparison} \label{sec:discussion}

We measure the stellar kinematics in the central region of M87, probing well within the SMBH sphere of influence and mapping the kinematics in exquisite detail. In this section, we place our results in context by comparing with previous multiscale stellar kinematic measurements of M87 obtained using ground-based facilities. 

\begin{figure*}
\begin{singlespace}
  \centering
  \includegraphics[width=\linewidth]{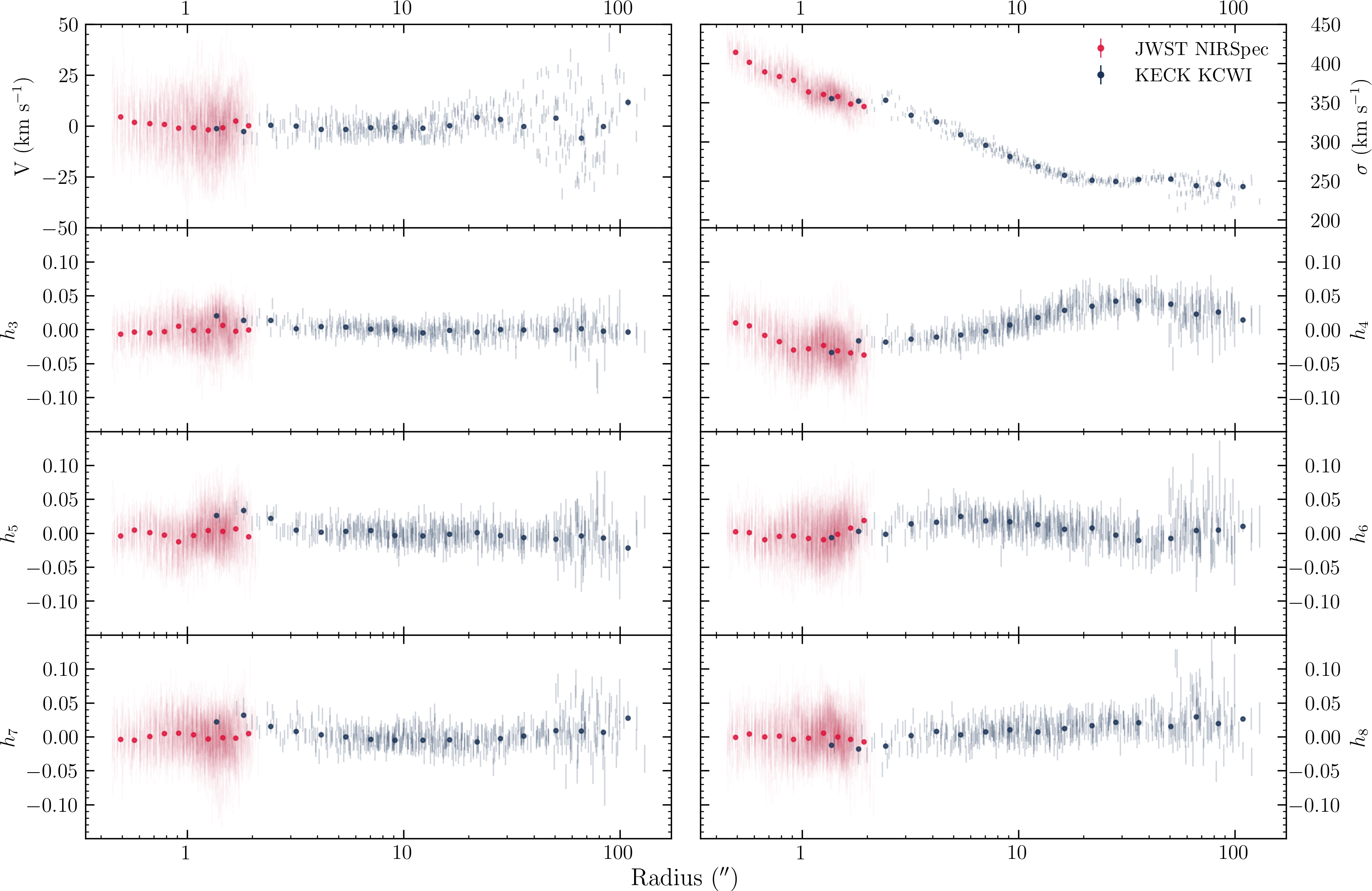}
    \caption{Juxtaposition of the small-scale NIRSpec measurements and the large-scale KCWI kinematics from \cite{Liepold2023}. The data have been folded and are plotted as a function of projected distance from the nucleus, hence multiple position angles are depicted for these IFU data. The radius is shown in logarithmic scale to encompass the wide spatial coverage of the KCWI dataset. The eight GH moments and uncertainties recovered from 2810 NIRspec spatial bins (red lines) transition nicely to those extracted from the 461 KCWI bins (black lines). For clarity, the red and black points are average kinematic values from NIRSpec and KCWI, respectively, within radial bins. The NIRSpec and KCWI kinematics shown are not symmetrized.}
  \label{fig:jwst_kcwi}
\end{singlespace}
\end{figure*}

\subsection{Large-Scale Measurements}

Since M87 has been the subject of numerous stellar-dynamical studies on large scales (e.g., \citealt{Dressler&Richstone1990, vanderMarel1994, Emsellem2004, Murphy_2011, Emsellem_2014, Sarzi_2018}), we focus on the most recent work \citep{Liepold2023} and refer the reader there for a comparison of prior wide-field kinematic measurements of M87. \cite{Liepold2023} used optical IFU data from Keck KCWI to extract stellar kinematics in 461 spatial bins from spectra with $S/N$ spanning $100-200$. They fit between $3900-5450$ \AA\ and report 8 GH moments over a 250$\arcsec \times$ 300$\arcsec$ FoV, using pPXF and template stars from the MILES library \citep{Sanchez-Blazquez2006, Falcon-Barroso2011} and an additive Legendre polynomial degree of 1 and a multiplicative degree of 15.

In Figure \ref{fig:jwst_kcwi}, we show $V-h_8$ from NIRSpec in all 2810 bins and from KCWI in 461 bins. To better see radial trends, we also present the average kinematics within bins of projected radius. As is the case for the NIRSpec kinematics, we show the unsymmetrized KCWI kinematics with (the small) offsets in the odd GH moments removed. We find that the small-scale NIRSpec kinematics nicely transition to the large-scale KCWI kinematics, including a continuation of the rapid rise in the velocity dispersion towards the center of the galaxy. After thorough examination of the NIRSpec stellar kinematics described in Section \ref{subsec:kintests}, no additional manual adjustments were needed to bring the two sets of kinematics into agreement. This is particularly noteworthy as the measurements were made from different telescopes/instruments, over distinct wavelength regimes, and using separate velocity template libraries.

While Figure \ref{fig:jwst_kcwi} highlights the radial trends in the M87 stellar kinematics, there are important structures in the full 2D kinematic maps. For example, the KCWI velocity field is twisted, and from 60\arcsec\ outward there is a 25 \kms\ rotational pattern with a kinematic axis that is 40$^\circ$ misaligned with the galaxy's photometric major axis. Consequently, \cite{Liepold2023} found that M87 has a strongly triaxial intrinsic shape. In a subsequent paper, we will combine the high-quality small-scale NIRSpec and large-scale KCWI data and constrain triaxial orbit-based models (e.g., \citealt{Quenneville2022}) over an impressive $\sim 40-12,000$ pc extent.

\subsection{Small-Scale Measurements}

Although there have been numerous large-scale stellar kinematic measurements made for M87 over the years, as well as gas-dynamical studies at the nucleus, extracting stellar kinematics at the center of M87 has proved challenging due to the galaxy's faint central surface brightness. There are only two prior high angular resolution studies of the stars within the central  $\sim 1\arcsec$ of M87, coming from \cite{Gebhardt2011} and \cite{Simon2024}, each employing different data sets and approaches. With AO Gemini NIFS data covering the first four $K$-band $^{12}$CO bandheads, \cite{Gebhardt2011} measured non-parametric LOSVDs in 40 spatial bins over the same FoV as NIRSpec. The spatial binning was done assuming a particular major axis position angle and yielded spectra with a $S/N =32-99$. \cite{Gebhardt2011} report that the inner Gaussian component of the NIFS PSF has a full width at half maximum (FWHM) of 0\farcs08 that contributes $\sim$40\% of the flux. Due to strong AGN contamination, nuclear spectra were discarded and the innermost kinematic measurement was made at $r \sim$ 0\farcs25.

\begin{figure}[h!]
\begin{singlespace}
  \centering
  \includegraphics[width=\linewidth]{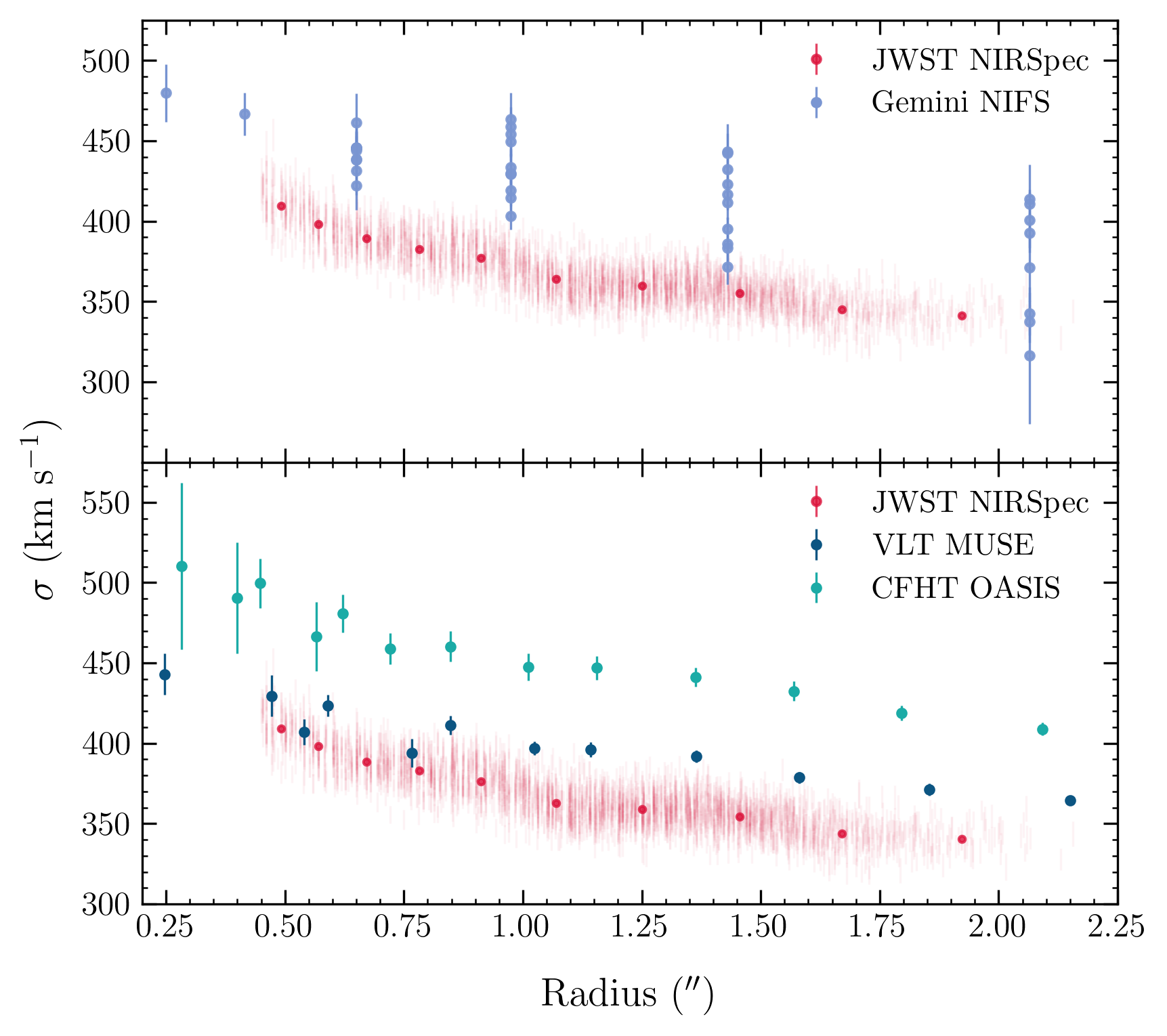}
     \caption{Comparison of the velocity dispersions measured from NIRSpec to those extracted from previous high angular resolution observations of M87. The data have been folded and multiple position angles from the IFU observations are displayed. The 2810 NIRSpec velocity dispersions with uncertainties (red lines) are shown when fitting four (top) and two (bottom) GH moments so that a fair comparison can be made to the prior measurements in the literature. The red points denote the average NIRSpec kinematic values within radial bins. \emph{Top:} The blue points with error bars are the velocity dispersions measured within 40 bins from AO NIFS data \citep{Gebhardt2011}. \emph{Bottom:} The velocity dispersions from AO NFM MUSE (navy) and good-seeing OASIS (teal) data, shown over the radial extent of NIRSpec, come from averages within radial bins from \cite{Simon2024}.}
   \label{fig:small_scale_sigma}
\end{singlespace}
\end{figure}

The top panel of Figure \ref{fig:small_scale_sigma} compares our NIRSpec velocity dispersions with those from NIFS. Our fidicual kinematics include measurements of $V - h_8$, however to properly compare to the NIFS data, we refit the NIRSpec spectra in the same way as described in Section \ref{sec:stellar_kin} but instead used four GH moments to parameterize the LOSVD. Likewise, \cite{Gebhardt2011} took their best-fit non-parametric LOSVDs, characterized them with a GH series up to $h_4$, and provided the moments in each of the spatial bins beyond $r \sim $ 0\farcs25 in their Table 2.

More recently, \cite{Simon2024} extracted stellar kinematics from new AO VLT MUSE observations with a PSF FWHM of 0\farcs049 and older good-seeing CFHT OASIS data with a PSF FWHM of 0\farcs561 (originally from \citealt{McDermid2006}). Both data sets cover optical wavelengths, $\sim 4800-5500$ \AA, and have a FoV of 7\farcs5$\times$7\farcs5 and 10\arcsec$\times$8\arcsec, respectively. \cite{Simon2024} spatially binned each data set to reach a $S/N \sim 50$ and assumed a Gaussian LOSVD with the $V$ fixed to the recessional velocity of M87. Thus, only $\sigma$ was measured with pPXF.

In the bottom panel of Figure \ref{fig:small_scale_sigma}, we show our NIRSpec kinematics along with those from MUSE and OASIS. To compare with \cite{Simon2024}, we refit the NIRSpec spectra following the methods in Section \ref{sec:stellar_kin}, but this time fit just $V$ and $\sigma$. \cite{Simon2024} found that the extracted $\sigma$ from both MUSE and OASIS was highly sensitive to the wavelength region fit and the additive polynomial degree used to adjust the template stars. In Figure \ref{fig:small_scale_sigma}, we plot the MUSE and OASIS points as shown in the bottom panel of Figure 5 in \cite{Simon2024}, which are their `RNI' degree $1$ kinematics averaged in radial bins. In addition, \cite{Simon2024} found their MUSE and OASIS velocity dispersions were systematically above the large-scale SAURON kinematics of M87 \citep{Emsellem2004, Cappellari_2011} over the common radial range of the data sets. To enforce a smooth transition to the SAURON kinematics, \cite{Simon2024} scaled their initial MUSE and OASIS velocity dispersions downward before running their main Jeans-based stellar-dynamical model. We note that we show the unscaled, original measurements of the MUSE and OASIS dispersions in Figure \ref{fig:small_scale_sigma}.

As can be seen in Figure \ref{fig:small_scale_sigma}, the velocity dispersion from NIFS is systematically above the NIRSpec dispersions at all radii by an average of $\sim$50 \kms, or $\sim$15\% of the NIRSpec value. Similarly, the velocity dispersions from MUSE and OASIS are systematically larger than the NIRSpec dispersions by an average of $\sim$30 \kms\ and $\sim$90 \kms, or $\sim$10\% and $\sim$25\% of the NIRSpec value, respectively, over the radial range covered by NIRSpec. The velocity dispersions from MUSE and OASIS, scaled by \cite{Simon2024} (not shown in Figure \ref{fig:small_scale_sigma}), are more similar to the NIRSpec values, generally falling at the upper end of the 1$\sigma$ uncertainties on the NIRSpec velocity dispersions. All four data sets (NIRSpec, NIFS, MUSE, and OASIS) exhibit a central rise in the velocity dispersion.

While the broad radial trends among the high-angular resolution data sets are analogous, the NIRSpec data provide a much improved kinematic view of the center of M87. With the NIRSpec data, there is a dramatic increase in the number of spatial bins, a substantial improvement in the $S/N$, and far more detailed characterization of the LOSVDs. JWST's unique combination of sharp angular resolution, enhanced sensitivity, and reduced background has opened a new window into the stellar dynamics at the centers of nearby massive galaxies with diffuse cores -- systems that were previously difficult to probe in detail.

\section{Conclusion} \label{sec:conclusion}

We present new stellar kinematic measurements of the central region of M87 using the JWST NIRSpec IFU with excellent $S/N$ over the $K$-band CO bandheads. These observations significantly improve upon prior ground-based high angular resolution data sets of M87 and provide detailed stellar kinematics in the immediate environment around the SMBH. 

We carefully examine the default JWST reduction pipeline, and implement a couple of modifications to enhance the data quality. In particular, we mask artifacts in each of the 16 dithered 2D calibrated images and we assemble the merged data cube in such a way that minimized extreme positive and negative artifacts, especially near the NRS1/NRS2 detector gap. Hence, the integrity of the spectra is maintained and robust stellar kinematics can be determined from the CO bandheads.

We conduct an in-depth exploration of the most appropriate way to extract the M87 stellar kinematics from the CO bandheads on the NRS1 detector. After exploring two empirical velocity template libraries (K2 OSIRIS and Winge GNIRS), to complement the main (synthetic) library we used (PHOENIX), we find consistent kinematics when masking \ion{Ca}{1} just blue-ward of the CO bandheads. The K2 OSIRIS and Winge GNIRS libraries are velocity template spectra that haven't yet been used for NIRSpec kinematic studies, but they are libraries that have been employed for near-infrared AO IFU data in the past. PHOENIX has been the library of choice for kinematic studies of local galaxies with NIRSpec. For M87, we noticed that the PHOENIX library struggled to fully reproduce the \ion{Ca}{1} feature and the even GH moments are biased low when \ion{Ca}{1} is included in the spectral fit. The K2 OSIRIS and Winge GNIRS libraries were able to match the observed \ion{Ca}{1}, and there is no difference in the inferred GH moments when the feature is fit versus when it is masked. Our final kinematics come from using the PHOENIX library with \ion{Ca}{1} masked.

In the case of M87, we find that the degree of the polynomial applied to the LOSVD-convolved template stars had an impact on the inferred even GH moments. With a larger fitting window covering four CO bandheads before the detector gap, a multiplicative polynomial of at least degree 3 was needed to prevent biased even moments. For the spatial bins with spectra exhibiting two or three CO bandheads before the detector gap, any degree ($1-6$) for the multiplicative polynomial returned consistent results. Since the wavelength range we fit over, and the number of CO bandheads before the detector gap, varied over the NIRSpec FoV, using too low of a polynomial degree yielded even GH moments that were too large in only a portion of the IFU, creating an artificial gradient. We adopted a degree 3 multiplicative polynomial to produce secure GH moments and kinematic maps with real spatial trends. Thus, when working with high $S/N$ NIRSpec data, we recommend using different template libraries with attention paid to the fit of the \ion{Ca}{1} line and there be an examination of the polynomial degree (or equivalently how the galaxy continuum is modeled). 

The M87 NIRSpec stellar kinematics transition seamlessly to the most recently published large-scale kinematics from KCWI, providing continuous coverage from the central parsecs to several kiloparsec scales. Together, we expect these datasets to enable tight constraints on the M87 \mbh, stellar mass-to-light ratio, dark matter halo, galaxy intrinsic shape, and orbital distribution throughout the galaxy and especially near the SMBH. Future work will apply the most general, triaxial stellar-dynamical models to the NIRSpec and KCWI kinematics, cementing M87 at the upper-end of the SMBH$-$galaxy relations and aiding in our understanding of SMBH and galaxy co-evolution.

\facilities{JWST(NIRSpec)}

\software{astropy \citep{astropy_2013, astropy_2018, astropy_2022}, JWST Science Calibration Pipeline \citep{jwstpipeline_v11.4_2023}, matplotlib \citep{hunter_2007}, numpy \citep{harris_2020}, pPXF \citep{Cappellari2023}, scipy \citep{virtanen_2020}, vorbin \citep{Cappellari2003}}

\begin{acknowledgments}

This work is based on observations made with the NASA/ESA/CSA JWST. The data were obtained from the Mikulski Archive for Space Telescopes at the Space Telescope Science Institute (STScI). The specific observations analyzed can be accessed via \dataset[https://doi.org/10.17909/vm4z-zk15]{https://doi.org/10.17909/vm4z-zk15}. STScI is operated by the Association of Universities for Research in Astronomy, Inc., under NASA contract NAS 5-03127 for JWST. These observations are associated with program GO-2228. Support for program GO-2228 was provided by NASA through a grant from the STScI, which is operated by the Association of Universities for Research in Astronomy, Inc., under NASA contract NAS 5-03127. E.~Liepold and C.-P.~Ma acknowledge support from the Heising-Simons Foundation. Research was conducted with the advanced computing resources provided by Texas A\&M High Performance Research Computing.

\end{acknowledgments}

\bibliography{M87_paper}

\end{document}